\documentstyle[aps,prl,epsf,floats]{revtex}

\begin{document}
\draft

%************************************************************
%***** Title ************************************************
%************************************************************

\twocolumn[\hsize\textwidth\columnwidth\hsize\csname@twocolumnfalse%
\endcsname
\title{Orbital Polarons in the Metal-Insulator Transition of Manganites}
\author{R.\ Kilian and G.\ Khaliullin}
\address{Max-Planck-Institut f\"ur Physik komplexer Systeme,
N\"othnitzer Strasse 38, D-01187 Dresden, Germany}
\date{15 March, 1999}
\maketitle

%************************************************************
%***** Abstract *********************************************
%************************************************************

\begin{abstract}
The metal-insulator transition in manganites is strongly influenced
by the concentration of holes present in the system. Based upon an
orbitally degenerate Mott-Hubbard model we analyze two possible 
localization scenarios to account for this doping dependence:
First, we rule out that the transition is initiated by a disorder-order
crossover in the orbital sector, showing that its effect on charge 
mobility is only small. Second, we introduce the idea of orbital polarons 
originating from a strong polarization of orbitals in the vicinity of holes. 
Considering this direct coupling between charge and orbital
degree of freedom in addition to lattice effects we are
able to explain well the phase diagram of manganites for
low and intermediate hole concentrations.
\end{abstract}

\pacs{PACS number(s): 72.80.Ga, 71.30.+h, 71.38.+i, 75.50.Cc}]

%************************************************************
%***** Body *************************************************
%************************************************************

The doping dependence of the properties of manganites, 
$R_{1-x}A_x$MnO$_3$ (where $R$ and $A$ represent rare-earth and 
divalent metal ions, respectively), poses some of the most
interesting open problems in the physics of these compounds.
First to be noticed is the peculiar asymmetry of the phase diagram that 
is most pronounced in the charge sector:
Regions of high ($x>0.5$) and low ($x<0.5$) concentration of holes are
characterized by such contrasting phenomena as 
charge ordering and metalicity, respectively. In the latter
region~-- which we wish to focus on~-- the metallic state
can be turned into an insulating one by  
raising the temperature above $T_C \approx 200$ - $350~\text{K}$.
Introducing the notion of double exchange \cite{ZAG} 
which associates the relative orientation of localized Mn $t_{2g}$ 
spins with the mobility of itinerant $e_g$ electron,
early work has identified this transition to be controlled 
by the loss of ferromagnetic order inherent to the metallic state.
It is believed that the lattice effects are also of crucial 
importance in this transition. Within the lattice polaron - double 
exchange picture \cite {MIL98}, the crossover from metallic to 
insulating behavior is controlled by the ratio of polaron binding
energy to the kinetic energy of charge carriers. Assuming the 
former quantity to be constant, a critical coupling strength leading 
to localization can be reached by reducing the kinetic energy, 
i.e., by raising temperature which acts via the double-exchange mechanism.
Spin disorder and spin polaron effects \cite {VARMA} further enhance
the carrier localization above $T_C$. The doping dependence of the 
metal-insulator transition, however, is not fully captured 
in this picture. Namely the complete breakdown 
of metalicity at hole concentrations below $x_{\text{crit}} \approx 
0.15$ - $0.2$ that occurs despite the fact that the fully saturated 
ferromagnetic state is sustained remains an open problem which we 
will address in this paper.

We analyze two mechanisms that could drive the localization 
of doped holes at small $x$. First, we explore the possibility 
of the metal-insulator transition to be controlled by a disorder-order 
crossover in the $e_g$-orbital sector. We outrule this scenario 
by showing that the mobility of doped holes depends only little 
on the degree of order in the orbital background. Second, we 
introduce the concept of orbital polarons in an orbitally degenerate 
Mott-Hubbard system~-- this constitutes the main idea of the paper. 
Similar to spin polarons in correlated spin systems such as cuprates, 
the orbital polaron is a natural consequence of strong correlations 
and the double degeneracy of on-site levels, the latter being a consequence
of the $e_g$-orbital degeneracy present in ferromagnetic manganites. 
We argue that the coupling between charge carriers and orbital degrees of 
freedom yields a strong polarization of $e_g$ orbitals in the neighborhood 
of holes that acts together with breathing-mode like lattice distortions 
to form combined orbital-lattice polarons. The effect is more 
pronounced at small $x$ because the polarizability of orbitals 
enhances as the characteristic energy scale of orbital fluctuations
$\propto xt$ reduces. Within the orbital-lattice polaron 
scenario we are able to explain well the phase diagram of manganites 
for low and intermediate doping levels.

{\it Orbital order-disorder and hole mobility}.~---
As the frustrating effect of holes is weakened at small $x$,  
a Jahn-Teller- and superexchange-driven transition from a disordered, 
strongly fluctuating orbital state to an orbitally ordered state 
is expected. This possibly suggests the localization process to be 
triggered by the establishment of order in the orbital sector,
which would naturally introduce a doping dependence due to the energy scale
$xt$ of orbital fluctuations. We analyze this possibility by comparing the 
mobility of holes moving in orbitally disordered and ordered backgrounds.
The relevant transfer Hamiltonian of $e_g$ electrons has to 
account for strong on-site Coulomb repulsions and the orbital degeneracy:
\begin{equation}
H_t = -\sum_{\langle ij \rangle_{\gamma}} \sum_{\alpha\beta} 
\left(t_{\gamma}^{\alpha \beta}
\hat{c}^{\dagger}_{i\alpha} \hat{c}_{j\beta}+\text{H.c.}\right)
\label{HT}
\end{equation}
with $\gamma\in\{x,y,z\}$. We use constrained operators 
$\hat{c}^{\dagger}_{i\alpha} = c^{\dagger}_{i\alpha}\left(1-n_i\right)$
that create electrons at site $i$ in orbital $\alpha$ only
under the condition that the site is empty. In orbital basis
$\alpha \in \{|3z^2-r^2\rangle, |x^2-y^2\rangle \}$ the 
intersite transfer matrices are:
\[
t_{x/y}^{\alpha\beta} = 
t\left(\begin{array}{cc}
1/4 & \mp\sqrt{3}/4\\
\mp\sqrt{3}/4 & 3/4
\end{array}\right), \quad
t_{z}^{\alpha\beta} = 
t\left(\begin{array}{cc}
1 & 0\\
0 & 0
\end{array}\right).
\]
Electron spins are treated within the double-exchange model and 
implicitly enter through the transfer amplitude $t$ only. To observe the 
strongly correlated nature of Hamiltonian (\ref{HT}) it is convenient to
introduce separate quasiparticles for charge and orbital degrees of 
freedom \cite{ISH97}. 
In the case of an orbitally disordered state we employ a slave-boson 
representation in which electrons are replaced by fermionic orbitons
$f_{i\alpha}^{\dagger}$ carrying orbital pseudospin and bosonic holons 
$b^{\dagger}_i$ carrying charge \cite{KIL98}.
On a mean-field level with parameters $\sum_{\alpha\beta} t^{\alpha\beta}_{\gamma} 
\langle f^{\dagger}_{i\alpha} f_{j\beta} \rangle_{\gamma} = \chi$ 
and $\langle b_i^{\dagger}\rangle = \sqrt{x}$ these quasiparticles can be 
decoupled. The characteristic energy scale of orbital fluctuations is then set
by the orbiton half band width $D_{\text{orb}} = 3xt$,
the mobility of holes by the holon half band width 
$D_{\text{hole}} = 3t$.

Orbital order in manganites is induced by a coupling of orbital pseudospins
mediated by, e.g., Jahn-Teller phonons and superexchange. The
corresponding Hamiltonian (see, e.g., \cite{KIL98,MAE97}) is of 
$XY$-type with internal frustration making it difficult to handle. To
simplify the discussion we simulate an orbitally ordered state by adding to 
Hamiltonian (\ref{HT}) the interaction term
\begin{equation}
H_J = -\frac{J}{2z} \sum_{\langle ij \rangle_{\gamma}} \sigma^x_i
\sigma^x_j e^{iq_{\gamma}}
\label{HJ}
\end{equation}
with $z=6$ and Pauli matrices $\sigma^x_i$. For $\bbox{q}=(\pi,\pi,0)$
Eq.\ (\ref{HJ}) favors a staggered 
$\left(|3z^2-r^2\rangle \pm |x^2-y^2\rangle\right)/\sqrt{2}$
ordering within $x$-$y$ planes repeating itself along the $z$ 
direction; this closely resembles the type of ordering observed 
experimentally in LaMnO$_3$ \cite{MUR98}.
The establishment of orbital order is indicated by a singularity in the 
static susceptibility $\langle \sigma^x\sigma^x\rangle_{\bbox{q}}$. In
random-phase approximation
\begin{equation}
\langle \sigma^x\sigma^x\rangle_{\bbox{q}} =
\frac{\langle \sigma^x\sigma^x\rangle^0_{\bbox{q}}}
{1+J_{\bbox{q}} \langle \sigma^x\sigma^x\rangle^0_{\bbox{q}}/2}
\label{RPA}
\end{equation}
with $J_q = J \left(\cos q_x + \cos q_y - \cos q_z \right)/3$.
Bare susceptibilities $\langle \cdots \rangle^0$ are treated employing 
the above mean-field description. Numerically solving for the
pole of Eq.\ (\ref{RPA}) with $J = 0.13~\text{eV}$ as estimated
from the structural phase transition observed in LaMnO$_3$ at 
$T = 780~\text{K}$ \cite{MUR98} and $t = 0.4~\text{eV}$
we find a critical doping concentration 
$x_{\text{crit}} = J/4t = 0.08$;
at concentrations below this critical value
an orbitally ordered state is to be expected.

We now turn to analyze the effect of orbital order onto the
mobility of holes. 
For this we study the evolution of the holon half band width,
which in the disordered state $D_{\text{hole}} = 3t$, as
orbital order develops.
Foremost, an important difference between 
models with orbital pseudospin and conventional spin is to be 
noticed: In the latter systems spin is conserved when electrons are 
transferred between sites. 
In contrast, the transfer Hamiltonian (\ref{HT}) of the orbital
model is nondiagonal in orbital pseudospin~--
the transfer matrices $t_x^{\alpha\beta}$, $t_y^{\alpha\beta}$,
and $t_z^{\alpha\beta}$ cannot be diagonalized simultaneously.
Hence there is a finite amplitude for holes to hop without 
perturbing the orbitally ordered background, leading to the formation of a 
wide coherent hole band; for the specific type of order introduced
above its half width is $D_{\text{hole}}^{\text{coh}} = 2t$.
This number seems to indicate a profound reduction in the mobility
of charge carriers by $\approx 30\%$ as compared to the disordered state.
It does, however, not reflect the full 
kinetic energy of holes which is further contributed to by incoherent 
processes involving the creation of excitations in the orbital
background. To study the influence of these processes we employ an 
``orbital-wave'' approximation that uses a bosonic description of orbital 
excitations of energy $J$ and a fermionic description 
of holons moving within a band of half width $2t$. Holons 
are coupled to orbital excitations via part of Hamiltonian (\ref{HT}).
In analogy to studies of spin systems \cite{KAN89}
we analyze the motion of a single hole employing a self-consistent Born 
approximation. Two limits can be treated analytically: As orbital excitation 
become very heavy, $J/t \rightarrow \infty$, the hole decouples from the 
orbital subsector. It then moves coherently due to the nondiagonality 
of transfer matrices alone, its motion being maximally 
constrained as $D_{\text{hole}} = 2t$. In the opposite limit, 
$J/t \rightarrow 0$, the hole moves completely incoherently in a cloud of 
an infinite number of orbital excitations suppressing orbital correlations 
in the background. The hole mobility then reaches 
its maximum $D_{\text{hole}} = 3t$ as in the disordered 
state, while the effective mass of the composite object formed of hole 
and excitation cloud becomes large, $m\rightarrow \infty$.
For a ratio $J/t = 0.3$ realistic for manganites one is closer to the 
latter limit: Numerically we find the establishment of orbital
order to reduce the kinetic energy of the hole by only less than $5\%$.
We therefore conclude that a disorder-order crossover in the orbital 
sector has only a secondary effect on the mobility of charge carriers, 
outruling it as a driving mechanism to initiate the metal-insulator transition.

{\it Orbital polarons }.~--- In the preceding paragraph we have 
considered charge carriers and orbitals to interact via the transfer
part (\ref{HT}) of the Hamiltonian. We will now show that in an orbitally
degenerate Mott-Hubbard system there also exists a direct 
coupling stemming from a strong polarization of $e_g$ orbitals on sites 
next to a hole. The lifting of orbital degeneracy is mediated by a 
displacement of oxygen atoms situated between the Mn sites as well as by 
the $e_g$ level splitting in the vicinity of positively charged holes. 
We estimate the magnitude of the degeneracy lifting, 
$\Delta = \Delta^{\text{ph}} + \Delta^{\text{ch}}$, as follows:
Holes respectively electrons are coupled to the lattice breathing
mode $Q_1$ and to two Jahn-Teller modes $Q_2$ and $Q_3$ by
\begin{eqnarray}
H_{\text{el-ph}} &=& \sum_i \Big(
-g_1 Q_{1,i} n^h_i + g_2 \left(Q_{2,i} \sigma^x_i + Q_{3,i} \sigma^z_i 
\right)\nonumber\\
&&+\frac{K}{2} \bbox{Q}_i^2 \Big),
\label{HLA}
\end{eqnarray}
where $n_i^h$ is the number operators for holes. Integrating over
oxygen displacements $\bbox{Q} = \left(Q_1,Q_2,Q_3\right)$ we obtain
$\Delta^{\text{ph}} = g_1 g_2 \sqrt{2}/(3K) \approx (g_1/g_2) E_{\text{JT}}$.
A lower bound for this quantity is given by the Jahn-Teller energy
$E_{\text{JT}} \approx 0.1$ - $0.2~\text{eV}$, assuming that coupling to 
the breathing mode is at least as strong as to the Jahn-Teller one.
On the other hand, the contribution to the $e_g$ level splitting
from the Coulomb interaction between the positively charged hole and 
an electron on a neighboring site is estimated taking into account the
covalency of Mn $3d$ and O $2p$ orbitals. It follows
$\Delta^{\text{ch}} \approx \case{3}{4}\gamma^2 e^2 R_{\text{Mn-Mn}}$, where the
covalency factor $\gamma = t_{pd}/\Delta_{pd}$ can be obtained
from the transfer amplitude and the charge gap between Mn and O sites,
$t_{pd} \approx 1.8~\text{eV}$ and $\Delta_{pd} \approx 4.5~\text{eV}$
\cite{SAI95}, respectively. 
With lattice spacing $R_{\text{Mn-Mn}} = 3.9~\text{\AA}$ this
leads to $\Delta^{\text{ch}} \approx 0.4~\text{eV}$.
In total, the polarization of $e_g$ levels on sites next to 
a hole yields an energy splitting $\Delta \approx 0.5$ - $0.6~\text{eV}$.
Being comparable in magnitude to the transfer amplitude $t$ this number
strongly indicates a direct coupling of charge and orbital
degrees of freedom to be of importance in manganites.

In the cubic system the Hamiltonian describing the above coupling is
\begin{equation}
H_{\text{ch-orb}} = -\Delta \sum_{\langle ij \rangle_{\gamma}}
n_i^h \tau_j^{\gamma},
\label{HPO}
\end{equation}
where $\tau_j^{\gamma} = \left( \sin\Theta \sigma^x + \cos\Theta 
\sigma^z \right)/2$ denotes orbital pseu\-dospin operators with angles 
$\Theta = \mp 2\pi/3,0$ corresponding to bonds in directions $\gamma = x,y,z$.
Hamiltonian (\ref{HPO}) promotes the formation of orbital polarons.
For low enough hole concentrations these consist of a bound state between
a central hole and the surrounding $e_g$ 
orbitals pointing towards the hole. Besides minimizing interaction energy
this configuration yields a large amplitude of virtual excursions of
$e_g$ electrons onto the empty site, thus lowering the kinetic
energy as well. In other words, the double exchange process is locally 
enhanced, providing a large effective spin of the orbital polaron.
This naturally explains the development of ferromagnetic clusters 
experimentally observed at temperature above $T_C$ \cite{TER97}.
At large doping concentrations, orbital polarons cannot be
considered as isolated anymore. In the presence of many holes, the
fast fluctuations of orbitals destroy the orbital-hole bound state. 
To elaborate on this point we calculate the orbital-hole 
binding energy $E_{\text{orb}}$ by considering a static hole in an 
orbital background. The upper bound for $E_{\text{orb}}$ is set by
the limiting case of uncorrelated, static orbitals: 
Hamiltonian (\ref{HPO}) yields $E_{\text{orb}} = 3\Delta(1-x)$
if all six orbitals surrounding a hole point towards the latter,
where the factor $(1-x)$ accounts for the fact that 
some of the orbitals might be empty. In the real system, however,
the binding energy is reduced by inter-orbital coupling with
energy scale $\propto J$ and orbital fluctuations of frequency
$\propto xt$ which are comprised in the effective ``stiffness'' of 
the orbital sector $D_{\text{orb}} = 3xt+J$.
Calculating the binding energy for a static hole in both orbitally 
ordered and disordered states we find an interpolating formula
\begin{equation}
E_{\text{orb}} = 3\left( \sqrt{\Delta^2(1-x)^2+D_{\text{orb}}^2}
-D_{\text{orb}}\right),
\label{Eb}
\end{equation}
which correctly reproduces the limits $\Delta\gg D_{\text{orb}}$ and 
$\Delta\ll D_{\text{orb}}$.
In marked contrast to conventional lattice polaron theory, the binding 
energy of the orbital polaron explicitly depends on 
the concentration of holes, foremost via the energy scale of orbital 
fluctuations $xt$: The more holes
are present in the system the faster orbitals fluctuate and the
smaller the binding energy $E_{\text{orb}}$ becomes. The
crossover from the small-polaron regime in which the orbital-hole
bound state is stable to the large-polaron regime in which the bound
state breaks up is governed by the dimensionless coupling 
constant $\lambda_{\text{orb}} = E_{\text{orb}}/D_{\text{hole}}$,
where $D_{\text{hole}}$ is the half band width of holes.
Since the lattice breathing mode in Eq.\ (\ref{HLA}) further
enhances the polaron formation, the total coupling constant 
$\lambda = \lambda_{\text{orb}} + \lambda_{\text{ph}}$ has an additional 
contribution $\lambda_{\text{ph}} = E_{\text{ph}}/D_{\text{hole}}$,
where $E_{\text{ph}}=g_1^2/(2K)$.
It is important to note that a critical value of 
$\lambda$ leading to localization can be reached 
either by increasing temperature which quenches $D_{\text{hole}}$ via 
double exchange or by lowering $x$ which enhances $E_{\text{orb}}$.

In the following we study in more detail the consequences of the 
polarization of orbitals around doped holes for the
phase diagram of manganites. We start from the strong-coupling
limit $\lambda \gg 1$ in which holes are substituted for by small polarons 
that move in a band whose width is reduced by an exponential factor
\begin{equation}
e^{-\eta} =  \exp\left[-\gamma\left(
\frac{E_{\text{orb}}}{D_{\text{orb}}}+\frac{E_{\text{ph}}}{\omega_0} 
\right)\right],
\label{ES}
\end{equation}
where $\omega_0$ is the frequency of dispersionless phonons. In deriving
Eq.\ (\ref{ES}) we neglect the interference between orbital
and lattice couplings. To simulate the crossover from the small polaron
to the free carrier picture we phenomenologically employ the function
$\gamma = \beta - \ln\left[\lambda\left(1+\beta\right)\right]/\lambda^2$
with $\beta = \left(1-1/\lambda^2\right)^{1/2}$ which has been proposed for 
strongly coupled phonon systems \cite{ALE94}; for $\lambda<1$ holes are assumed
to move freely as $\gamma = 0$. A feedback of Eq.\ (\ref{ES}) into the binding 
energy, Eq.\ (\ref{Eb}), resulting from the quenching of orbital fluctuations
$\propto xt e^{-\eta}$ as small polarons begin to form sharpens the 
transition. Next we turn to the coupling between the charge and spin sector 
through the double-exchange mechanism. On the one hand, the transfer amplitude
$t = t_0\left[\left(1+m^2\right)/2\right]^{1/2}$ depends on the magnetization 
$m$. This is determined from $m = B_S(\alpha m)$ with $\alpha=3S/(1+S)T_C/T$, 
where $B_S$ is the Brillouin function and $S=\case{3}{2}+(1-x)\case{1}{2}$ the
average spin of Mn $3d$ electrons per site.
On the other hand, charge transfer mediates a ferromagnetic
interaction between sites. The effective exchange coupling constant is
\begin{equation}
J_{\text{eff}} = \frac{t\chi}{2S^2} \left[ x(1-x)e^{-\eta}
+x_0(1-x)^2\right]
\label{JEF}
\end{equation}
with $x_0 = 2t\chi/U$, where $U$ is the on-site repulsion, and
mean-field parameter $\chi \approx \case{1}{2}$. The Curie temperature
is obtained from $T_C = \case{1}{3}\nu z S(S+1)J_{\text{eff}}$;
the fitting parameter $\nu$ compensates for 
the overestimation of $T_C$ in mean-field treatment.
The first term in squared brackets of Eq.\ (\ref{JEF}) stems from the 
coherent motion of polarons and represents the conventional double-exchange
contribution to $T_C$. The second term is due to 
superexchange. It is insensitive to the localization process and dominates 
ferromagnetic interaction at low doping levels \cite{MAZ98,END98,KHA99}.

\begin{figure}
\noindent
\centering
\epsfxsize=0.9\linewidth
\epsffile{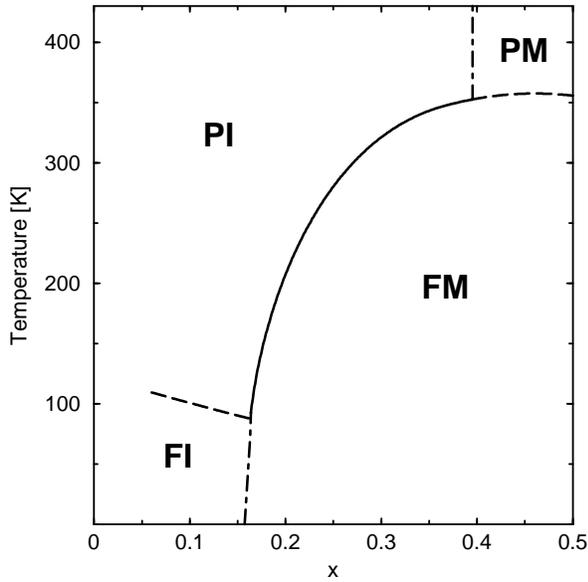}
\caption{Magnetic and electronic phase diagram obtained within
the present theory. Dashed and 
dashed-dotted lines represent magnetic and electronic transitions, 
respectively, the simultaneous transition in both channels is denoted by 
a solid line. The phases are: paramagnetic insulator (PI), paramagnetic 
metal (PM), ferromagnetic metal (FM), and ferromagnetic
insulator (FI).}
\label{FIG1}
\end{figure}

To illustrate the interplay between the system of equations 
(\ref{Eb}) - (\ref{JEF}) we numerically extract from them the $T$-$x$ phase 
diagram. The following parameters are chosen:
The orbital polarization energy is set to $\Delta = 0.5~\text{eV}$, yielding
a binding energy comparable to the phononic one
$E_{\text{ph}} = 0.6~\text{eV}$; the phonon frequency is 
$\omega_0 = 0.05~\text{eV}$, the interaction between orbitals 
$J = 0.13~\text{eV}$, the bare transfer amplitude $t_0 = 0.4~\text{eV}$,
and $x_0 = 0.1$. 
The fitting parameter $\nu=0.5$ \cite{RUS74} is adjusted to reproduce
the values of $T_C$ observed for La$_{1-x}$Sr$_x$MnO$_3$ \cite{URU95}. 
The result is shown in Fig.\ \ref{FIG1}. 
Due to the doping dependence of the orbital polaron binding energy the 
system is more insulating at low and more metallic at high doping levels.
This is manifested in the absence of a metallic phase at doping levels
$x<0.15$ and the appearance of a paramagnetic metallic phase at
$x>0.4$. In the region in between a simultaneous magnetic and electronic 
transition from a ferromagnetic metal to a paramagnetic insulator occurs. 
We note that the loss of charge mobility triggers static orbital order.
A long-range orbitally ordered state has in fact been experimentally 
detected in the insulating regions of La$_{0.88}$Sr$_{0.12}$MnO$_3$ 
\cite{END98}. In general, however, the ordered state is expected to 
have orbital and Jahn-Teller glass features due to the presence of 
quenched orbital polarons. 
The orbital polaron provides a strong local ferromagnetic coupling between 
the planes in c-direction, thus explaining the existence of the isotropic 
ferromagnetism even in the insulating phase at small doping. Finally it is 
worth to notice that the phase diagram in this theory is highly
sensitive to the transfer amplitude $t_0$ as this parameter enters in
the polaron binding energy, Eq.\ (\ref{HPO}). 

We conclude that orbital polarons are an intrinsic part
of an orbitally degenerate Mott-Hubbard system and play an
important role in the physics of manganites. Being comparable
in magnitude to lattice effects this new mechanism naturally 
introduces a doping dependence into the polaron binding energy 
via the degree of orbital fluctuations. Accounting
for both orbital and lattice effects we are able to reproduce
well the important features of the phase diagram of manganites.

%************************************************************
%***** Acknowledgement **************************************
%************************************************************

%************************************************************
%***** References *******************************************
%************************************************************

\end{document}